\documentclass[useAMS,usenatbib, usegraphicx]{mn2e}
\usepackage{amssymb}
  \def\apjs{ApJS} 
\def\apjl{ApJ}   \def\aap{A\&A}
   
\let\ga=\gtrsim \let\la=\lesssim

\title[Growth of Red Galaxies] 
{The Growth of Red Sequence Galaxies in a Cosmological Hydrodynamic Simulation}

\author[Gabor et al.]{
J. M. Gabor,$^{1,2}$\thanks{Email:jgabor@as.arizona.edu}
R. Dav\'e$^{2}$ 
\\ $^{1}$CEA-Saclay, IRFU, SAp, F-91191 Gif-sur-Yvette, France\\
   $^{2}$University of Arizona, 933 N. Cherry Ave, Tucson, AZ 85721\\
}

\begin{document}


\pagerange{\pageref{firstpage}--\pageref{lastpage}} \pubyear{2010}

\maketitle
\label{firstpage}

\begin{abstract}
  We examine the cosmic growth of the red sequence in a cosmological
  hydrodynamic simulation that includes a heuristic prescription for
  quenching star formation that yields a realistic passive galaxy
  population today.  In this prescription, halos dominated by hot gas
  are continually heated to prevent their coronae from fueling new
  star formation.  Hot coronae primarily form in halos above $\sim
  10^{12}M_{\sun}$, so that galaxies with stellar masses $\sim
  10^{10.5}M_{\sun}$ are the first to be quenched and move onto the
  red sequence at $z>2$.  The red sequence is concurrently populated
  at low masses by satellite galaxies in large halos that are starved
  of new fuel, resulting in a dip in passive galaxy number densities
  around $\sim10^{10}M_{\sun}$.  Stellar mass growth continues for
  galaxies even after joining the red sequence, primarily through
  minor mergers with a typical mass ratio $\sim$ 1:5.  For the most
  massive systems, the size growth implied by the distribution of
  merger mass ratios is typically $\sim 2 \times $ the corresponding
  mass growth, consistent with observations.  This model reproduces
  mass-density and colour-density trends in the local universe, with
  essentially no evolution to $z=1$, with the hint that such relations
  may be washed out by $z\sim 2$.  Simulated galaxies are increasingly
  likely to be red at high masses or high local overdensities.  In our
  model, the presence of surrounding hot gas drives the trends both
  with mass and environment.

\end{abstract}
\begin{keywords}
galaxies:evolution -- galaxies:formation
 \end{keywords}

\section{Introduction} 

A substantial fraction of galaxies today inhabit a tight locus in
colour-magnitude space known as the red sequence.  Such galaxies typically
have elliptical morphologies, little or no star formation, little
cold gas, and live in dense environments.  They host the majority of
stellar mass in the local universe \citep{hogg02_redgals}, and therefore
are a critical population for understanding the global evolution of
galaxies across cosmic time.

Despite a growing wealth of data, the origin of red sequence galaxies
is still not fully understood.  In the currently-favored cold dark
matter (CDM) paradigm, massive galaxies should be surrounded by halos
of hot gas that can potentially cool and fuel new star
formation~\citep{white91}.
Yet these galaxies show little cold gas or
ongoing star formation.  Debate continues as to what physical
mechanisms can halt star-formation and keep it halted for much of
cosmic time, with recent work favoring feedback from active galactic
nuclei (AGN) as the primary energy
source~\citep[e.g.][]{croton06,hopkins08_ellipticals}.  

Deep surveys have begun to place constraints on bright passive
galaxies over cosmic timescales, out to $z>2$ \citep[e.g.][]{bell04,
  faber07, brown07, cool08, kriek08, stutz08, taylor09, brammer09,
  marchesini10, whitaker10}, and recent surveys should probe down to
stellar masses $\sim10^9 M_{\sun}$ at such redshifts
\citep[e.g. CANDELS;][]{grogin11,koekemoer11}.  Current data indicates
that the red sequence has grown by $\times 2$ from $z\sim 1$ to 0, but
the most massive galaxies were already in place 7+ Gyrs ago.
Furthermore, high-resolution imaging has revealed that passive
galaxies are typically much smaller at $z>1$, by factors $\sim 5$
compared to their present-day descendents \citep{daddi05, vandokkum08,
  vandokkum10}.  These data provide strong constraints on
passive galaxy formation models, and only recently have such
constraints been explored in the context of hierarchical models
\citep{khochfar06, fan08, nipoti09a, nipoti12, naab09, oser10,
  shankar2010, oser12, cimatti12}.

Numerous physical processes are thought to contribute to the formation
and evolution of passive galaxies, and many potential effects have
proven difficult to rule out.  Gas-rich galaxy mergers that induce a
starburst and rapid black hole growth may drive the initial quenching
of massive red galaxies \citep{springel05_mergers_ellipticals,
  hopkins06_redgals, hopkins08_ellipticals}.  Alternatively, hot
gaseous coronae that form in halos above $\sim 10^{12}M_{\sun}$,
supplemented by additional heating from an active galactic nucleus
(AGN) or another energy source, may starve galaxies of fuel for
star-formation \citep{birnboim03, keres05, dekel06, croton06,
  cattaneo06}.  Among satellite galaxies additional processes may
drive quenching: starvation and/or ram-pressure stripping by the hot
intergalactic medium, or glancing interactions between galaxies
\citep{gunn72, abadi99, quilis00, larson80, bekki02, richstone76,
  moore98}.  All these processes may act in combination to form
passive galaxies as observed.

Once ``red and dead,'' the evolution of these galaxies may be
deceptively complex.  Although there is little or no new star
formation, the stellar population sheds $\sim 1/2$ of its mass via
stellar winds \citep[][]{jungwiert01, bc03}.  Mergers
can alter galaxy structure\citep[e.g.][]{barnes90}, change the
characteristic size of the galaxy \citep{cox06_kinematics}, strip
stars into the ICM \citep{gallagher72, murante04}, and possibly add
cold gas.  Such processes may all contribute to the observed evolution
of the red sequence.


In this work, we study passive galaxy growth from high redshift until
today using cosmological hydrodynamic simulations.  These simulations
incorporate physically-motivated but heuristic quenching mechanisms
\citep[described in][]{gabor11} that yield a $z=0$ population of passive
galaxies whose colour and luminosity distributions match observations.
In particular, we include a prescription where we ensure that halos
dominated by hot gas are continuously heated to keep circum-galactic
gas hot.  This simple and extreme prescription, approximating the
effects of a heat source such as an AGN radio jet, successfully cuts
off the fuel supply for star formation in massive halos which eventually
yields passive galaxies.  
Although significant challenges remain for our favoured model, it
provides a general qualitative guide for models where hot massive
halos are the main drivers of passive galaxy formation.  While
\citet{gabor11} focused on the $z=0$ red galaxy population and
compared various quenching mechanisms, here we focus on our most
successful quenching mechanism and its implications for passive galaxy
evolution from high redshift until today.  Our model should not be
considered physically correct in detail, but rather as broadly
illustrative of the impact of some (unspecified) quenching mechanism
that keeps hot halo gas hot on passive galaxy evolution over the past
10 billion years.

We begin by presenting an overview of our simulations in \S \ref{sec.sims}.
Then we present our results, beginning in
\S \ref{sec.evol} where we examine
number density and colour evolution, 
\S~\ref{sec.growth} where we look at how passive galaxies grow
in mass and size via mergers, and \S~\ref{sec.envir} where we study mass and colour evolution versus
environment.
Finally, we summarize and discuss our conclusions in \S~\ref{sec.summary}.


\section{Simulations} 
\label{sec.sims} 

\subsection{Simulation methodology} 

We analyze the same cosmological hydrodynamic simulations described in
\citet{gabor11}.  For completeness, we summarise the simulations here.
They were run with an extended version of the N-body + smoothed particle
hydrodynamics code GADGET-2 \citep{springel05}.  In addition to the basic
N-body and hydrodynamics calculations, our version of the code includes
sub-resolution modelling of gas cooling, star-formation, a model for
chemical enrichment via AGB stars and both Type Ia and core-collapse
supernovae, galactic winds associated with star-formation, and simple
prescriptions for quenching star-formation.

We include both primordial and metal-line cooling assuming collisional
ionisation equilibrium \citep{sutherland93}, with the metallicities
self-consistently tracked within simulations; see
\citet{oppenheimer06,oppenheimer08} for details.  We include heating
from a metagalactic photo-ionising background~\citep{haardt01},
assuming all particles are optically thin.  


For star formation, we employ the two-phase model of
\citet{springel03} based on the analytic description of
\citet{mckee77}.  Gas particles above a density threshold of
0.13~cm$^{-3}$ are treated as cold, star-forming clouds embedded
within a hot diffuse medium, and they form stars on a timescale
consistent with the observed relation to surface gas density
\citep{kennicutt98}.  Star-forming gas particles are converted into
collisionless star particles stochastically, with a probability
derived from the star formation rate.

As a gas particle is undergoing star formation, it self-enriches with
metals from core-collapse supernovae.  Furthermore, star particles
share energy, mass, and metals with neighboring gas particles as a
result of stellar mass loss from AGB stars and Type Ia supernovae.
Stellar mass loss is calculated by assuming a \citet{chabrier03}
initial mass function, applying the stellar population models of
\citet[][ hereafter BC03]{bc03}, which account the mass lost by a
stellar population at discrete times after the initial star-formation
event.  We use Type Ia supernova rates from \citet{scannapieco06}, and
each such supernova results in the production of metals (mainly iron)
that are shared with neighboring gas particles.


Our code further includes feedback in the form of galactic winds
driven by star-formation \citep{oppenheimer06, oppenheimer08}.  Just
as a star-forming gas particle has some probability of being converted
into a star particle, it has a probability of being kicked in a wind,
which is given by $\eta$ (the mass loading factor) times the star
formation probability.  A wind particle is expelled from its host
galaxy at a velocity $v_w$ typically a few hundred km s$^{-1}$.
The
velocities are chosen to match observations of local galaxy winds
\citep{martin05,rupke05} and those seen at higher
redshifts~\citep[e.g.][]{steidel10}.  In particular, $v_w \propto$ the galaxy velocity dispersion, and $\eta \propto$ the galaxy circular
velocity as calculated from an on-the-fly galaxy
finder~\citep{oppenheimer08}; these scalings are predicted for
momentum-driven winds \citep{murray05}.  A galaxy in a $10^{12}
M_{\sun}$ halo at $z=1$ typically expels a wind with $v_w \approx
500$~km~s$^{-1}$ and $\eta \approx 1.7$ \citep{oppenheimer10}.
Once launched, winds are
decoupled from the hydrodynamic calculation until they reach a density
10\% of the critical density for star-formation, up to a maximum
duration of 20~kpc~$v_w^{-1}$. With this prescription, our
simulations match a broad array of observational constraints on
star-forming galaxies and the intergalactic medium
\citep{oppenheimer06, oppenheimer08, oppenheimer10, dave06, dave07,
  finlator07, finlator08}.

Despite their broad success, winds driven by star-formation generally
do not result in the formation of red and dead galaxies.  We
incorporate an additional heuristic quenching model specifically to
solve this problem \citep{gabor11}.  For this model, we run a
spherical overdensity algorithm on-the-fly to identify galaxy halos
and distinguish gas above and below 250,000 Kelvin in each galaxy's
halo.  In halos with $>60$\% of all gas above this temperature cutoff,
we apply constant heating (at every time-step) to all circum-galactic
halo gas to force it to remain at the halo virial temperature.  We
exclude star-forming gas particles from this heating.  Such heating
could plausibly result from an AGN radio jet, but our model in fact
uses more energy than thought to be available from observed AGN
sources \citep{gabor11}.  Other possible heat sources include cosmic rays
\citep[e.g.][]{mathews09} or graviational heating by infalling gas
clumps \citep{dekel08, khochfar08, birnboim11}.  We remain agnostic
about the exact mechanism, and instead examine whether an effective
heating source can lead to a realistic population of quenched
galaxies.  We emphasize that our simulations do not track the growth
of supermassive black holes.


The heating we apply to circum-galactic gas is sufficient to prevent
it from condensing onto the galaxy, thus starving the galaxy of new
fuel for star-formation.  Over $1-2$~Gyr, star-formation will exhaust
any remaining cold gas, and star-formation will cease.  This model
results in a bimodal colour distribution of galaxies, and red galaxy
luminosity functions that match observations of the local universe.
Although the model has some difficulties (like the required energy
mentioned above), it should be a good representation of models where
galaxies are quenched due to starvation enabled by their surrounding
hot coronae.

\subsection{Runs and galaxy identification}

For this paper, we use a simulation of a $48 h^{-1}$ comoving Mpc
random cosmological cube with 256$^3$ dark and 256$^3$ gas particles
that incorporates all the above physics.  We use a \emph{Wilkinson
  Microwave Anisotropy Probe} concordance cosmology \citep{komatsu09}
with $H_0 \equiv 100h = 70$km s$^{-1}$ Mpc$^{-1}$, matter density
$\Omega_m=0.28$, baryon density $\Omega_b=0.046$, a cosmological
constant with $\Omega_{\Lambda} = 0.72$, root mean square mass
fluctuation at separations of 8~Mpc $\sigma_8 = 0.82$, and a spectral
index of $n=0.96$.  Our simulation uses a gravitational softening
length of 3.75$ h^{-1}$ kpc.  The initial gas particle mass is $1.2
\times 10^8 M_{\sun}$, the typical star particle mass is half that,
and our simulation results in $\sim 3000$ resolved galaxies at $z=0$.
We have also used a simulation with $2 \times 384^3$ particles and the
same volume to explicitly check that our main results are
resolution-converged, as expected from previous work
\citep{finlator06, dave11a}.  We note that high-resolution zoom
simulations of individual galaxies can produce quenched galaxies
without explicit feedback, but they have too much stellar mass
\citep{naab07}.

We save snapshots of the simulation at 108 redshifts, starting at $z=30$
and ending at $z=0$.  The time between snapshots ranges from a few tens
of Myrs at high redshift to $\sim 300$ Myr at low redshift.  The snapshots
contain information for simulation particles, such as position, velocity,
mass, metallicity, gas density, gas temperature, and star-formation rate.
From these particle data, we determine galaxy properties to compare
with observables.

We use  {\tt Skid}
\footnote{http://www-hpcc.astro.washington.edu/tools/skid.html}
to identify galaxies \citep[cf. ][]{gelb94, keres05}.  {\tt Skid}
provides a list of member particles (star and star-forming gas) for each
simulated galaxy.  The sum of member star particle masses is then the
galaxy stellar mass, and the instantaneous star formation rates of the
gas particles are summed to give the star formation rate of the galaxy.

We then calculate galaxy spectra using the models of \citet{bc03}, as
in \citet{finlator06}.  We treat each star particle as a single
stellar population with an age and metallicity determined directly in
the simulation.  By adding up the spectra of all star particles within
a galaxy, we obtain the spectrum of that galaxy, from which we measure
galaxy colours and magnitudes in various bands.  Since we focus on
the passive galaxy population that is thought to be mostly dust-free
\citep{lauer05}, we ignore dust reddening and extinction.

\subsection{Building merger trees}

Beyond knowing galaxy properties at a given redshift, we wish to study
the histories of individual galaxies as they evolve.  For this we must
connect each galaxy at $z=0$ to its progenitor galaxies at earlier
redshifts: i.e., build a merger tree.  We do so by determining, for
every star particle in a given galaxy, which galaxy from an earlier
snapshot that star particle lived in.  If the star formed recently it
wouldn't have lived in any previous galaxy.  By tracing star particles
in this way, we determine the immediately preceding progenitors of
each galaxy.  Galaxies with multiple progenitors must have undergone a
merger (or at least some galaxy interaction, such as stellar
stripping).

Creation of a merger tree from a cosmological simulation involves
handling a number of pathological circumstances owing to the stochastic
nature of the gas and star particles.  A common annoyance is that if
two galaxies fly by each other in a close encounter, {\tt Skid} (and most
other galaxy finders) will often identify them as a single galaxy
during one or a few timesteps, but then again as separate galaxies at
later times.  Sometimes star particles are not assigned to any group.
While these issues do not affect the majority of
galaxies and hence do not strongly impact the overall results, they must
nevertheless be handled in an appropriate manner.

To surmount these difficulties, we follow strategies described in
\citet{maller06}, who dealt with the same issues.  If a star particle
is not assigned to any {\tt Skid} groups at a given timestep, we trace its
history through earlier timesteps to find the last galaxy it was in,
and assign it to the descendant of that galaxy.  When two {\tt Skid} groups
at one timestep have a single common progenitor at an earlier
timestep, we assume that the two groups are a merging pair and assign
them to the same galaxy.


The above prescription assumes that once galaxies are joined by {\tt
  Skid}, they will eventually merge.  But some fly-by interactions
will not lead to mergers.  To account for this, we identify galaxies
at $z=0$ that are composed of multiple {\tt Skid} groups and separate
those groups into distinct galaxies.  We find the earliest timestep
when both groups' stars are assigned to the same galaxy, and from then
on assign a new galaxy ID to those stars in the smaller of the two
groups.  Once every star particle is assigned to a galaxy at every
timestep, we have the full merger history.

\subsection{Environment measures}
In order to study \emph{where} red galaxies emerge in our simulation,
we measure the local galaxy density around {\tt Skid} galaxies.  We
use two density measures.  An intuitive and simple approach is to
count the number of simulated galaxies within $1h^{-1}$~Mpc of the
galaxy of interest~\citep{blanton06}.  For some purposes this measure
is too noisy, as it places galaxies into discrete bins of density, and
some galaxies have few neighbors within $1h^{-1}$~Mpc.  Thus, as an
alternative we use {\tt Smooth}
\footnote{http://www-hpcc.astro.washington.edu/tools/smooth.html},
which calculates the mean density at each galaxy's location after
smoothing over the nearest 16 galaxies with a spline kernel (similar
to an SPH density calculation).  This allows us to quantify low
densities less stochastically than by simply counting galaxies within
some fixed distance.  In some cases we use a galaxy's local
overdensity measured
relative to the mean resolved galaxy density in our simulation volume.
Based on numerous tests, our density estimator based on {\tt Smooth}
is comparable to and correlated with common observable density
estimators such as a 5th-nearest neighbor density
\citep[e.g.][]{kovac10}. 

\section{Evolution in colour and number density}
\label{sec.evol}
\subsection{Color-mass diagrams and stellar mass functions}
\label{sec.numdens}
\begin{figure*}
\includegraphics[width=168mm]{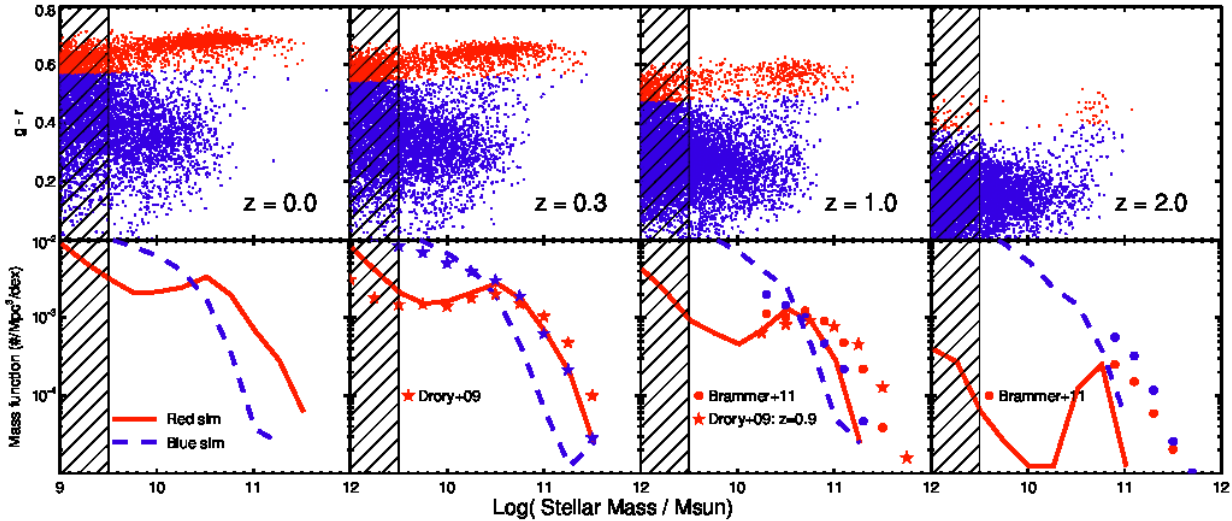}
\caption{Redshift evolution of colour-mass diagrams (top row) and
  galaxy stellar mass functions (bottom row).  Redshift increases from
  left to right, as labeled.  Red and blue galaxies are separated
  using a redshift-dependent cut, as illustrated by the red and blue
  colour-coding in the CMD.  In the mass function panels, solid lines
  denote the simulated red sequence, dashed lines the simulated blue
  cloud, and observed data points are taken from the literature.  The
  hatched region indicates poorly resolved galaxies.  Massive red
  galaxies are present at $z=2$, but not in sufficient numbers 
  compared to observations. There is a dearth of intermediate
  mass galaxies ($\sim10^{10} M_{\sun}$) that is more pronounced at
  higher redshifts.}
\label{fig.colormass}
\end{figure*}

In this section we examine how colours and number
densities (parameterized via the galaxy stellar mass function) evolve
with redshift in our simulations.  We note that our simulations do not
perfectly match present-day observations of passive galaxies, as
detailed in \citet{gabor11}.  For instance, the colors are slightly
too blue and the red sequence does not have the correct slope, 
owing to an under-production of massive galaxies' metallicities by
  $\sim 50$\% \citep[see ][ for a full discussion]{gabor10, gabor11}.
Nonetheless, our simulations clearly identify two general trends in
passive galaxy evolution.  First, passive galaxies emerging at
$z>2$ are at the massive end of the galaxy stellar mass function,
followed shortly by small satellites, then finally the intermediate region
of $10^{10} M_{\sun}$ galaxies.  Second, the most massive red galaxies in
our model grow substantially at late times.

Figure \ref{fig.colormass} shows colour-mass diagrams and galaxy
stellar mass functions for simulated galaxies at redshifts 0, 0.3, 1,
and 2.  The $g-r$ colours are rest-frame SDSS filter bands, and do not
include any correction for obscuration due to dust -- these are
intrinsic colours.  Dust reddening can induce large changes in galaxy
colors \citep[see][]{gabor10}, but for a given simulated galaxy the
dust reddening is highly uncertain.  We thus focus on the more robust
intrinsic properties of simulated galaxies.

Our quenching model leads to an obvious bimodality in (intrinsic)
galaxy colors, which corresponds to a bimodality between star-forming
and quiescent galaxies.  We divide the galaxy population into blue and
red using a straight line in $g-r$ colour versus $r$ magnitude space
in the same way as in \citet{gabor11}.  We evolve the normalization
(but not the slope) of our colour separation in a simple linear way
that approximately accounts for passive evolution: $y_{\rm sep}(z) =
y_{\rm sep,z=0} - 0.1 z$.  Here $y_{\rm sep}$ is the y-intercept
(i.e. colour) of the line of separation, and $z$ is the redshift.  The
hatched region on the left of the figure denotes the mass range of
galaxies that may be inadequately resolved.  The colour-mass diagrams
show strong growth in the passive galaxy population from
$z=2\rightarrow 0$.  Both the blue and red galaxy populations become
redder with time, at roughly the same rate.


The lower set of panels of Figure \ref{fig.colormass} shows galaxy
stellar mass functions separated into blue (dashed line) and red
(solid line) populations.  We plot mass function observations taken
from \citet{drory09} and \citet{brammer11}, who likewise split
galaxies into star-forming and quiescent populations (blue vs. red
points).  Since dust reddening adds difficulty and complexity to the
separation of star-forming and quiescent galaxies (and since we use
different separation criteria), we emphasize that comparisons with
observations are only qualitative.  We note that \citet{brammer11}
paid particular attention to separating truly passive from dusty
star-forming galaxies.

The simulated blue galaxy stellar mass function evolves only weakly
with redshift, but the red galaxy mass function undergoes drastic
growth from $z>2$ to 0.  The buildup of passive galaxies
reflects the quenching of star-forming galaxies due to starvation in
our model (see \citet{gabor11}).  The red sequence grows from mostly
non-existent at $z\sim 3$ to dominating the galaxy population at
$z\sim0$.  The mass functions clearly show that the most massive
galaxies are the first to become red, followed by low-mass galaxies
that turn out to be satellites to the massive ones.  This directly
results from the fact that the most massive galaxies in our
simulations are the first to form a corona of hot gas.  Massive
galaxies have massive dark matter halos, and only massive dark matter halos
form stable hot gas coronae~\citep{birnboim03}.


\subsection{A dearth of passive galaxies at $M_*\sim 10^{10} M_{\sun}$}

 Examination of Figure \ref{fig.colormass} reveals a
  characteristic ``dip'' in the simulated red galaxy mass function at
  $M_* \approx 10^{10} M_{\sun}$, which persists at all redshifts but
  is more prominent at high redshifts.  The red mass function peaks at
  high masses $M_* \approx 10^{10.6} M_{\sun}$, declines to
  intermediate masses $M_* \approx 10^{10.0} M_{\sun}$, and then rises
  again to low masses $M_* \lesssim 10^{9.8} M_{\sun}$.

The first passive galaxies form at the massive end by $z\approx3$
(which is an underestimate since our limited simulation volume does
not probe the most massive structures at $z>3$).  Soon thereafter,
small satellites of the first quenched galaxies become red as well,
owing to starvation of gas in growing hot halos~\citep{simha09}.
Intermediate-mass galaxies with $M_{\rm stellar} \approx 10^{10}
M_{\sun}$ are the last to quench, filling in the gap between the
centrals and satellites such that by $z\sim 0$ this bimodality within
the red sequence is not so evident.  Therefore, a strong prediction of
this quenching mechanism is that there is a distinct gap between the
massive and low-mass ends of the red sequence that becomes more
pronounced with redshift.

A variety of observational studies already suggest that there are
fewer red galaxies at intermediate-mass than at the high-mass peak,
especially at high redshift, as our simulated mass functions imply
\citep{stutz08, drory09, mortlock11, rudnick09, yang09, rudnick12,
  tal12}.  At $z\approx 0$, recent studies show a slight decline in
the red galaxy mass function from high to intermediate masses,
requiring a characteristic double-Schecter function to get a good fit
\citep{yang09, peng10, baldry08, baldry12}.  Unlike our simulation,
however, observations do not typically show an upturn toward lower
masses \citep[but see][]{drory09, yang09}.  Some semi-analytic models
show an analogous mass bimodality for elliptical galaxies
\citep{delucia12}, although the driving processes are likely
different.

Although further observations will be required to confirm that there
are more high-mass than intermediate-mass quenched galaxies
(especially at higher-redshift), this trend naturally emerges from our
quenching model.
Our simple model, where hot gas dictates the
formation of the red sequence, provides a compelling physical picture,
which we describe next.

\subsection{Central and satellite evolution along the red sequence}\label{sec.centralsat}

The separation in mass range between central and satellite galaxies
along the red sequence is more pronounced at earlier epochs, leading
to a sort of bimodality along the red sequence itself.  This is
naturally understood within a hierarchical structure formation
model, as we outline here.

In our model, the initial emergence of the massive red sequence is
directly tied to the requirement that a galaxy's halo be dominated by
hot gas for quenching to begin.  Galaxy evolution models have long
predicted that gas in the haloes of galaxies should be virialised and
thus hot \citep{rees77, silk77, white78}.  Gas in haloes with masses
below a few $\times 10^{11} M_{\sun}$, however, will have short
cooling times, so that it never reaches the virial temperature of the
halo~\citep{binney77,white91,birnboim03, keres05}.  In our simulations
with metal-line cooling the threshold for a hot gas-dominated halo is
$\sim 10^{12} M_{\sun}$ \citep{gabor10}, with little or no redshift evolution.

The key assumption in our model is that in such hot-gas dominated
halos, gas is very inefficiently deposited onto the galaxy, likely
owing to the contribution of some preventive feedback process(es)
like AGN heating~\citep[e.g.][]{dekel06, croton06, cattaneo06, bower06,
birnboim07}.  Since galaxy stellar masses are well-correlated with halo
masses below the cutoff for quenching, only central galaxies with the
highest stellar masses will live in hot halos at early times.

As the universe evolves and individual halos grow, more galaxy halos become massive enough to
support hot coronae.  Massive central galaxies thus continue to
move onto the red sequence.  Our hot gas fraction cutoff of 60\%
roughly corresponds to a stellar mass of about $10^{10.5} M_{\sun}$,
though there is significant scatter.  Once galaxies exceed this
stellar mass, they will tend to migrate toward the red sequence
over the span of $1-2$~Gyr.  We note that this mass cutoff is
unlikely to be universal at $z\ga 2$, since there are seen to be a
number of star-forming galaxies with stellar masses approaching
and even exceeding $10^{11} M_\odot$~\citep[e.g.][]{daddi07}.  These galaxies may be fed by cold streams that are dense enough to penetrate the hot halo \citep{dekel09}.

What happens with satellite galaxies?  When a central galaxy becomes
quenched by its own hot corona, its satellites are likely to be
embedded within that hot corona as well.  Based on our criteria for
quenching, these satellites will be quenched as well.  In LCDM, the
most massive subhalo is typically less than 10\% of the mass of its
parent halo \citep{gao04} -- when a central galaxy of mass $10^{10.5}
M_{\sun}$ is quenched, its satellites will usually be $<10^{9.5}
M_{\sun}$.  This creates a ``gap'' where there are no red galaxies
around $10^{10}M_{\sun}$.

As time progresses, massive quenched halos merge with nearby halos
owing to hierarchical growth.  For example, a star-forming galaxy
  with a mass of $M_*=10^{10} M_{\sun}$ may fall into a more massive,
  group-sized halo.  As the smaller halo falls into the region
dominated by hot gas, its central galaxy becomes a satellite and is
quenched by the hot ICM on a $1-2$~Gyr timescale~\citep{simha09,
  skibba09_satellites, wetzel11}.  These infalling (former) central
galaxies begin to populate the intermediate-mass portion of the red
sequence.  Therefore hierarchical growth combined with starvation by
the hot corona results in a gradual ``filling in" of the
intermediate-mass red sequence.  These trends will be explored in more
detail, particularly in relation to environment, in \S\ref{sec.envir}.

%
%

In summary, our model predicts that galaxies
with $M_{\rm stellar}\sim 5\times10^{10}M_{\sun}$ are the first to
become passive.  Galaxies in the ``gap" at $M_{\rm
stellar}\sim 10^{10}M_{\sun}$ catch up with time.  The key feature
driving these trends is that our quenching mechanism, tied to the
hot gas coronae that form in halos $\sim 10^{12}M_{\sun}$, selects
a characteristic mass for passive galaxies.  That is, once a
star-forming galaxy achieves a mass $\gtrsim10^{10.5} M_{\sun}$,
it will become quenched and move to the red sequence.  Less massive
galaxies are only quenched as satellites (or nearby halos) of more
massive halos.  More massive red galaxies can only be obtained via
merging.  While quantitative details may vary with parameter choices,
these are generic outcomes for hierarchical quenching models keyed
to a critical halo mass.

\subsection{Paths to the red sequence}

\begin{figure}
\includegraphics[width=84mm]{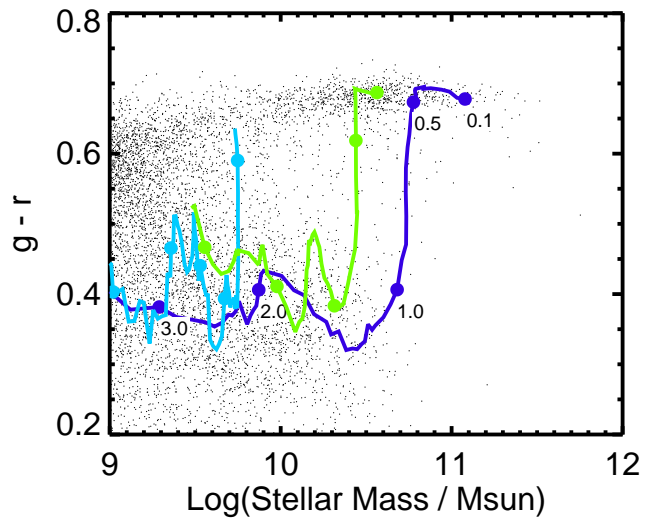}
\caption{Example evolutionary paths in the colour-mass diagram.  We
show the paths of three galaxies (coloured lines) as they evolve
through the colour-mass diagram, as represented by $z=0$ galaxies
(small points) in the background.  This is analogous to the schematic
diagram of \citet{faber07}.  Dots indicate redshifts along
the evolutionary paths, as labelled for the most massive galaxy.
Galaxies grow in stellar mass along the blue sequence, then move
vertically to the red after quenching.  Once on the red sequence,
galaxies grow only via mergers and accretion of satellites.  Such
growth can be substantial, as for the path of the most massive galaxy
shown.}
\label{fig.cmd_paths}
\end{figure}

Here we focus on the path of individual galaxies in colour-mass
space: Given a red galaxy at $z\sim0$, what path did it take to get
there?  It turns out that massive galaxies in our models follow
paths much like those in the schematic diagram of \citet{faber07}:
they move onto the red sequence at masses near the turnover in the
mass function, then grow further through mergers and accretion of
satellite galaxies.

In Figure \ref{fig.cmd_paths} we show the paths of three typical
galaxies (coloured lines) in the colour-mass diagram.  We plot a
background of simulated galaxies at $z=0$ (points), and we correct
for passive evolution of colours for the galaxy paths in the same
way that we evolved our line separating blue and red galaxies \S
\ref{sec.numdens}.  We do not include dust reddening.

All three of these galaxies build stellar mass via star-formation for
several Gyrs before moving onto the red sequence.  Once quenched, they
stop building stellar mass and cross vertically and rapidly over the
green valley to the red sequence in $\la 2$~Gyr~\citep[see the
  star-formation histories in ][]{gabor11}.  The least massive galaxy
makes this transition at late times ($z\sim 0.1$), and does not change
in mass or colour after reaching the red sequence because it does not
have time to undergo mergers.  The most massive galaxy, in contrast,
reaches the red sequence at $z\sim 0.5$ when its mass is several
~$\times 10^{10} M_{\sun}$, and then continues to grow in mass along
the red sequence via mergers, eventually obtaining a mass of $\sim
10^{11} M_{\sun}$.

More broadly, these paths are fairly representative for galaxies in their
respective mass ranges.  Low-mass galaxies tend to move onto the red
sequence at late times, and grow little after doing so, while massive
galaxies become passive at earlier epochs and grow in mass by merging
with smaller galaxies.

\subsection{Specific star formation rates}
\label{sec.ssfr}

\begin{figure*}
\includegraphics[width=7in]{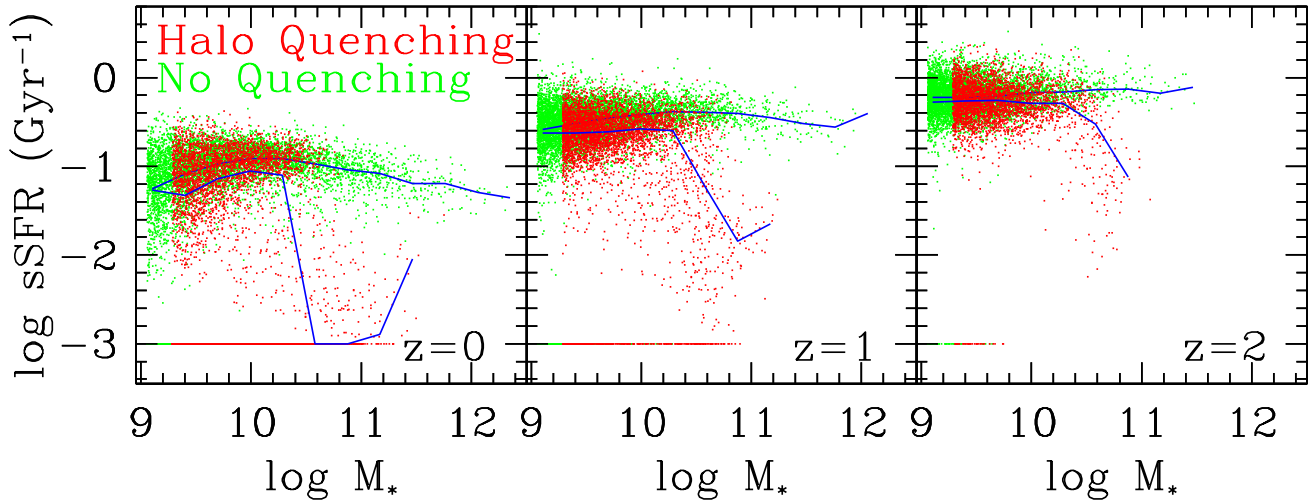}
\caption{Specific star formation rate (sSFR) as a function of stellar
  mass at redshifts 0 (left), 1 (middle), and 2 (right).  Red points
  correspond to our halo quenching model and green points to a
  simulation without quenching.  Blue solid lines show the running
  median sSFR for both models.  Many quenched galaxies have no
  star-formation, so we assign a minimum sSFR of $10^{-3}$~Gyr$^{-1}$.
  Quenching has little impact on the sSFR-$M_*$ relation below
  $\sim10^{10.5} M_{\sun}$, but causes a drop in sSFR above this
  threshold at all redshifts.  The transition quenching mass
  is blurred by satellite galaxies $\sim10^{10}M_{\sun}$ becoming quenched.}
\label{fig.sfrmstar}
\end{figure*}

A key observational constraint on the transition of galaxies onto
the red sequence is provided by the distribution of specific star
formation rates (sSFR) versus stellar mass.  This contains similar
information to the colour-mass plots above, but here we additionally
compare our current quenching simulations with previous similar
runs without quenching from \citet{dave11a}.  We particularly examine
the stellar mass range over which the transition to passive occurs,
since the observed lack of a sharp delineation in stellar mass
between passive and star-forming galaxies has been used as an
argument against a quenching prescription based on halo mass (as
ours is, approximately).

Figure~\ref{fig.sfrmstar} shows sSFR as a function of stellar mass
for a simulation without any quenching (green points) and the halo
quenching run presented in this paper (red).  Blue lines show a
running median for each.  Galaxies with sSFR$\leq 10^{-3}$~Gyr$^{-1}$
are shown along bottom of the plot.  The no-quenching run employs
the same cosmology, volume, input physics (besides quenching), and
outflow model, but it has somewhat better resolution ($\sim \times 3$).


The form and evolution of the sSFR for the non-quenched case is
discussed extensively in \citet{dave11a}.  Key points are that the
relation is flat at high-$z$ moving towards a mildly negative slope
by $z\sim 0$, sSFR increases with redshift at a given stellar mass
steadily out to $z=2$, and the scatter is typically $\sim 0.2$~dex
or less.  The relation fundamentally arises from gravitationally-driven
smooth cold accretion that dominates galaxy fueling, and diminishes
with time as the Universe expands~\citep[e.g.][]{dave08b, cen11}.

The impact of halo quenching is clearly seen at masses $\ga 3\times
10^{10}M_\odot$, the same mass where many galaxy properties are
observed to undergo a strong transition~\citep{kauffmann04}.  Below
this transition mass, there is a minimal difference ($<0.2$ dex) in
the quenched and non-quenched relations.  Above this transition mass,
the sSFR drops quickly towards small values, which is qualitatively
similar to observations~\citep[e.g.][]{salim07}.  The transition mass
evolves very little with redshift, being only marginally higher at
$z\sim 2$.  The highest mass galaxies are still forming a ``frosting''
of stars at a much reduced yet non-trivial rate
\citep{trager98,trager08}, reflecting the infall of satellites that
are not fully quenched.
In our model there are very few
galaxies above the transition mass on the main star-forming sequence, in
conflict with observations.  This conflict is also found in SAMs with
similar quenching mechanisms \citep{somerville08}.

Particularly noteworthy are the intermediate-mass galaxies that we
argued earlier are ``filling in" the gap of the red sequence at
stellar masses of $\sim 10^{10}M_\odot$.  
As these galaxies become more numerous towards lower redshifts, the
transition between quenched and unquenched is significantly blurred,
such that many quenched galaxies are present at masses below the
nominal transition mass.  Hence the argument that the observed
gradual transition from unquenched to quenched galaxies rules out
a model based on halo mass quenching (which our model approximates)
may not be justified.  While our simulations qualitatively resemble
observations, it is worth noting that such environmental quenching
processes are not as robustly modeled as we would like in our
simulations, and hence a more quantitative comparison will have to
await improved modeling of the interactions of galaxies moving
through hot halo gas.

\section{Growth along the red sequence}
\label{sec.growth}
\subsection{Mergers since quenching}
\begin{figure*}
\includegraphics[]{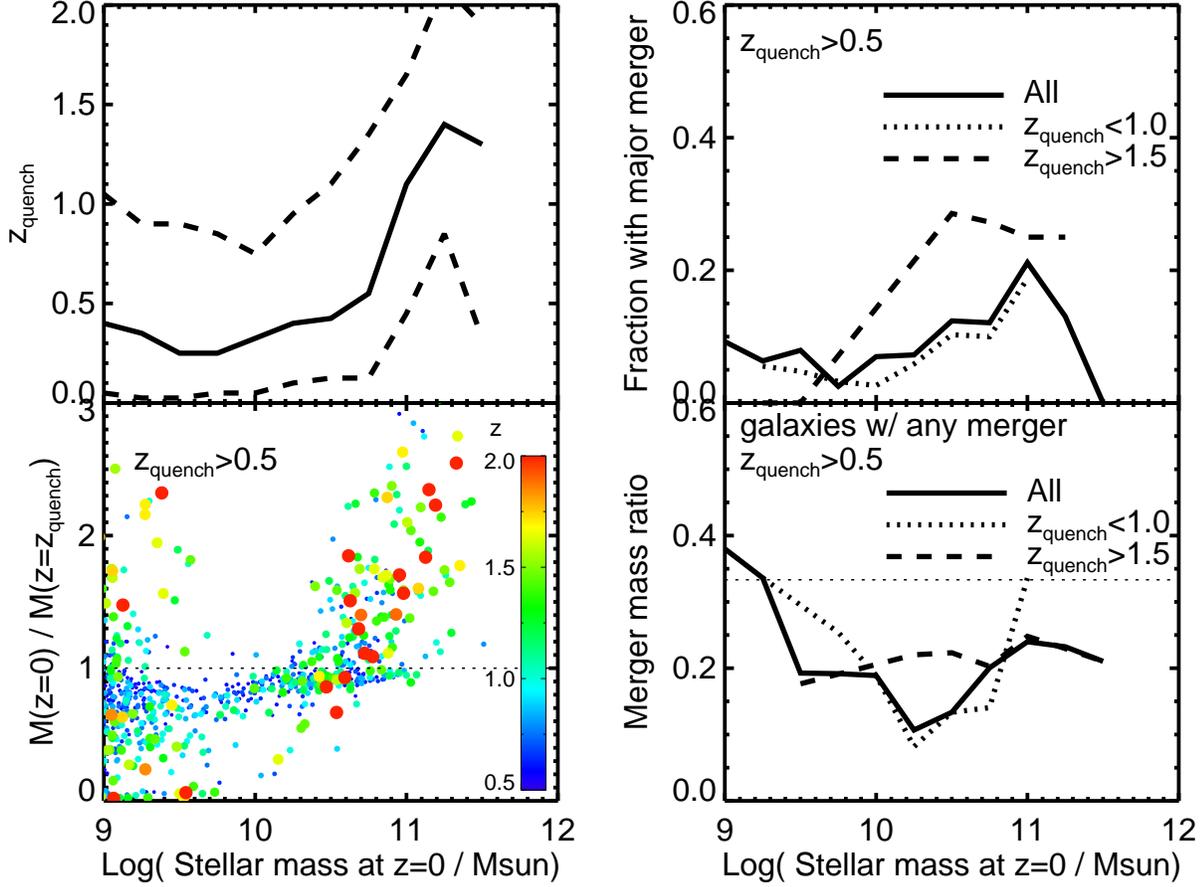}
\caption{Growth of galaxies along the red sequence.  In our model, the
  most massive galaxies typically grew by factors of a few since being
  quenched at $z\sim 1$, and that mass growth is dominated by minor
  mergers. 
{\bf Top Left:} median (solid line), 10th and 90th
  percentiles (dashed) of the redshift at which red sequence galaxies
  were quenched ($z_{\rm quench}$) in different mass bins.  
{\bf  Bottom Left:} fractional mass growth of red sequence galaxies
  since $z_{\rm quench}$, as a function of stellar mass at $z=0$,
  among galaxies quenched before $z=0.5$.  Galaxy points are
  colour-coded by $z_{\rm quench}$, and galaxies with higher $z_{\rm
    quench}$ have larger symbol sizes.  
{\bf Top Right}: the fraction
  of galaxies per bin that have undergone at least one major merger
  since being quenched, among galaxies with $z_{\rm quench} > 0.5$.
  The fraction for galaxies with $0.5 < z_{\rm quench}< 1.0$ (dotted)
  is typically lower than that for galaxies quenched earlier,
  $z_{\rm quench} > 1.5$ (dashed).  
{\bf Bottom Right}: the mass-weighted
  average merger mass ratio, among galaxies with $z_{\rm quench} >
  0.5$.  The mean merger is always a minor merger ($<1/3$, thin dotted
  line).  Galaxies quenched at high redshifts (dashed line) all have a
  merger mass ratio of $\approx 0.2$, while those quenched at somewhat
  lower redshifts (dotted) show stronger variation with mass.
}
\label{fig.mergers}
\end{figure*}

We have seen that a galaxy does not stop growing and evolving once
it reaches the red sequence (see Figure~\ref{fig.cmd_paths}).  Its
existing stellar population ages, that population sheds mass via
stellar winds, and it obtains new stars (and gas) via mergers.  Here
we show that the most massive galaxies, which are quenched at early
times ($z>1$), often grow by factors of a few between quenching and
the present day.  In contrast, low-mass red galaxies (mostly
satellites) tend to lose mass through stellar evolution and stripping
without undergoing significant mass growth via mergers.

The top left panel of Figure \ref{fig.mergers} shows the distribution
of the redshift at which galaxies are quenched, $z_{\rm quench}$,
as a function of the final stellar mass at $z=0$.  We plot the 50th
(solid), 10th, and 90th percentiles (dashed) of the distribution
in each bin.  The quenching redshift is defined as the redshift
when a galaxy first appears red in the colour-magnitude diagram,
with an evolving dividing line between red and blue galaxies as
described in \S\ref{sec.numdens}.  Low-mass galaxies on the $z=0$
red sequence, which are predominantly satellites, are quenched at
fairly late times (generally $z<1$).  Massive galaxies
$\gtrsim10^{11}M_{\sun}$ are quenched earlier, typically at $z\ga
1$ and beyond.

The bottom left panel of Figure \ref{fig.mergers} shows, as a
function of the $z=0$ stellar mass, the fractional mass growth of
red sequence galaxies between the redshift of quenching and the
present day, $y\equiv M_{\rm stellar}(z=0) / M_{\rm stellar}(z=z_{\rm
quenched})$.  Since many galaxies are quenched at late times, and
such galaxies have not had time to undergo mergers, we only plot
the mass growth for galaxies with $z_{\rm quench} > 0.5$.  This
gives a long enough time baseline ($\sim 5$~ Gyr, much longer than
the halo dynamical times) to assess red galaxy growth via mergers.
Galaxy points are colour-coded by $z_{\rm quench}$, as indicated
on the right side of the panel.

Departures from the $y=1$ line (i.e. no change in stellar mass, dotted
line) increase with quenching redshift: The blue points that are the
most recently quenched galaxies typically lie near $y=1$, and in fact
are slightly below owing to stellar mass loss.  At higher quenching
redshifts, galaxies span a wider range in $y$.  In some rare cases,
galaxies lose mass catastrophically by tidal interactions with other
galaxies, and end up near $y=0$.  A significant proportion of galaxies
quenched at earlier times with $M>10^{10.5} M_{\sun}$, however, have
grown substantially, by factors up to $\sim 3$.  Since these massive
galaxies live in high-density environments, and they've been quenched
for longer, they have had more opportunities to acquire mass via
mergers than their lower-mass counterparts.  Hence $y$ reflects the
amount of mass growth by mergers in massive galaxies since the time of
quenching, and in fact underestimates the growth since stellar mass
loss also occurs.  We note that a small number of galaxies become
quenched at $z<0.7$ yet undergo mass growth by $> \times2$ via several
major mergers.  Some low-mass galaxies with $y>2$ are only marginally
resolved in our simulation.


The top right panel of Figure \ref{fig.mergers} shows the fraction of
$z=0$ red sequence galaxies that have undergone at least one major
merger (1:3 mass ratio or larger) since being quenched.
Post-quenching \emph{major} mergers are generally rare, happening in
typically only $\sim 10\%$ of galaxies, and up to $\sim 20\%$ of
galaxies with $M_*\sim 10^{11} M_{\sun}$.  Galaxies quenched at the
earliest times have major merger fractions as high as 30\%.
Semi-analytic models indicate major merger fractions
reaching $>50$\% for massive galaxies~\citep{delucia06, wang08}.
These studies, unlike ours, have considered the integrated merger
history of the model galaxies.  Since mergers are more common at
higher redshifts \citep[e.g. ][]{hopkins10_mergers}, restricting
ourselves to mergers since the time of quenching naturally leads to a
lower fraction of galaxies with at least one merger.  In our model,
major mergers play a sub-dominant role in the overall growth of
massive red galaxies.

With little or no in situ star formation, the growth of red galaxies
must therefore owe predominantly to minor mergers.  The bottom right
panel of Figure \ref{fig.mergers} shows the mean merger mass ratio as
a function of $z=0$ stellar mass.  By weighting each merger by its
mass, we can assess the contribution that mergers of a given mass
ratio make to the overall growth of passive galaxies.  The typical
merger mass ratio for red sequence galaxies in our simulations is
$\sim 0.2$ or $1:5$, below the commonly-used $1:3$ cutoff for major
mergers (thin dotted line).  This typical value agrees with that
obtained in higher-resolution individual galaxy re-simulations by
\citet{oser12}.  At all stellar masses, red galaxy mass growth is
dominated by minor mergers.

In summary, massive red sequence galaxies in our simulation grow by
factors of up to several in mass between the time they are quenched
(typically $z\sim 1$) and redshift 0.   Lower mass passive galaxies
(both satellites and centrals) tend to be quenched at later epochs,
and grow less, typically even diminishing in mass owing to stellar
evolution.  Mass growth at all masses is dominated by minor mergers
with a characteristic ratio of 1:5, with at most a 30\% contribution
from major mergers.

\subsection{Size evolution}
\begin{figure}
\includegraphics[width=84mm]{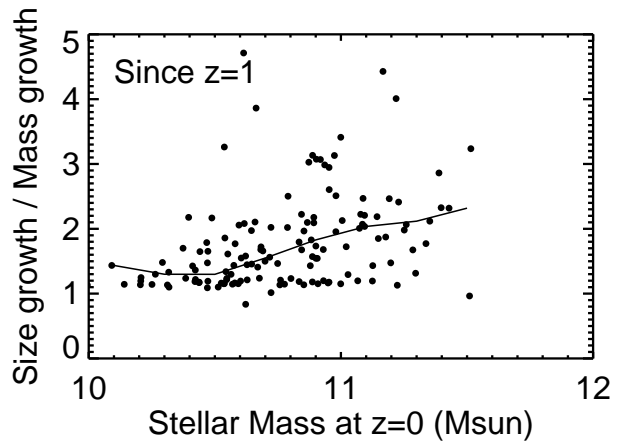}
\caption{The ratio of size growth to mass growth for red sequence
  galaxies, since $z=1$, as a function of their stellar mass at $z=0$.
  A solid line indicates the median.  More massive galaxies show more
  size growth than lower-mass ones.  In some rare cases the galaxy
  size may evolve several times ($\sim 3$) more than the mass.  This
  is roughly in line with observations, where the ratio is typically
  $\sim 2$ for evolution since $z\sim2$.}
\label{fig.size}
\end{figure}

Observations indicate that massive passive galaxies are more compact
at high redshifts than at $z\sim 0$ by factors of at least 2 and possibly up to 6
\citep{toft07, vanderwel08, vandokkum08, trujillo07, trujillo11}.  While observational biases
presented an initial concern, recent work suggests that much of the
evolution must be physical, and probably arises from a combination of
several physical processes.  These processes include adiabatic expansion
associated with stellar mass loss, major mergers, minor mergers, and
evolution in mass-to-light ratios \citep[e.g.][]{boylin-kolchin06,
naab09, bezanson09, hopkins10}.

Our simulations make specific predictions about the distribution of
merger mass ratios that contribute to red galaxy growth.  Therefore,
even though the spatial resolution of our simulations is not well-suited
to study the internal structure of individual galaxies, we can use simple
analytic merger-based estimates to approximate the size growth that our
simulated galaxies might undergo.

We construct a simple model for the effects of mergers following the
formalism presented in \citet{naab09}, which arises directly from the
virial theorem plus energy conservation.  The model is appropriate for a
purely collisionless system, so we implicitly ignore any effects of
dissipation -- since this model concerns quenched galaxies this is a
reasonable simplification, but it may lead to an over-estimate of the
size growth due to mergers.  Given a stellar accretion event (i.e. a
merger) of mass $M_a$ onto a galaxy of mass $M_g$, the ratio of final
to initial size of the galaxy is given by:
\begin{equation}
\frac{r_f}{r_i} = \frac{(1+\eta)^2}{1 + \eta \epsilon}.
\end{equation}
Here, $\eta = M_a / M_g$ is the fractional mass increase, and $\epsilon =
<v_a^2> / <v_g^2>$ is the ratio of the two galaxies' stellar populations'
mean square velocity dispersions.  Using high-resolution simulations,
\citet{naab09} has shown that this formula accurately describes the size
evolution of early-type galaxies undergoing hierarchical mergers.

To estimate size growth, we must therefore estimate the mass ratio and
the velocity dispersion ratio between the satellite (i.e. less
massive) and central (more massive) galaxies.  We know the ratio of
stellar masses directly from our merger tree.  For the velocity
dispersion, since we do not accurately resolve it in our simulations
\citep{oppenheimer08}, we instead employ an observational result
relating it to stellar mass from \citet{gallazzi06}: $\log \sigma =
-0.895 + 0.286 \log M_{\rm stellar}$.  With this relation, we obtain
$\epsilon = (\sigma_a / \sigma_g)^2 = (M_a / M_g)^{0.57}$, which gives
comparable results to the theoretical relation with an exponent of
$2/3$ at any redshift \citep{mo98}.  Thus, for each merger event, we
can calculate the fractional size growth of the remnant galaxy.
Taking each $z=0$ red galaxy, we consider its entire merger history
since being quenched.  We account for the size growth of each of its
merger events and multiply them to find the galaxy's total fractional
size growth (although we cannot determine whether the galaxies start
out small as observed).

We show the results of this exercise in Figure \ref{fig.size}, which
shows the size growth relative to mass growth ($r_f M_i/r_i M_f$) as a
function of stellar mass.  We plot this quantitiy instead of merely
size growth because massive red galaxies in our simulations may not
grow the correct amount of stellar mass, as hinted by the mass
functions in \S\ref{sec.numdens}.  The size growth relative to the
mass growth of galaxies is a reflection of the prevalence of
minor vs. major mergers.

Based on this simple modeling, minor mergers appear to be numerous enough
to drive factors of several in size growth of passive galaxies over
time.  Galaxies with stellar masses $>10^{11} M_{\sun}$ show typical
size growth a factor of 2 greater than their mass growth, indicating
that a series of minor mergers tends to puff up these galaxies, as argued by
\citet{naab09}.  Lower mass galaxies have not grown as much since moving
onto the red sequence, since as we showed in Figure~\ref{fig.mergers}
they undergo less merging.  These results, using a simple analytic model,
are broadly consistent with the idea that minor mergers drive the rapid
size growth seen in massive red galaxies since high redshift.

In summary, a simple model of red galaxy size growth driven by
mergers shows that massive galaxies with $\ga10^{11} M_{\sun}$ in
our simulation typically have grown in size by a factor of 2 relative
to their mass growth since $z=1$.  The degree of size growth increases
with stellar mass.  These results are broadly consistent with observations
as well as previous simulation results.

\section{Environmental Factors}
\label{sec.envir}
\begin{figure*}
\includegraphics[width=168mm]{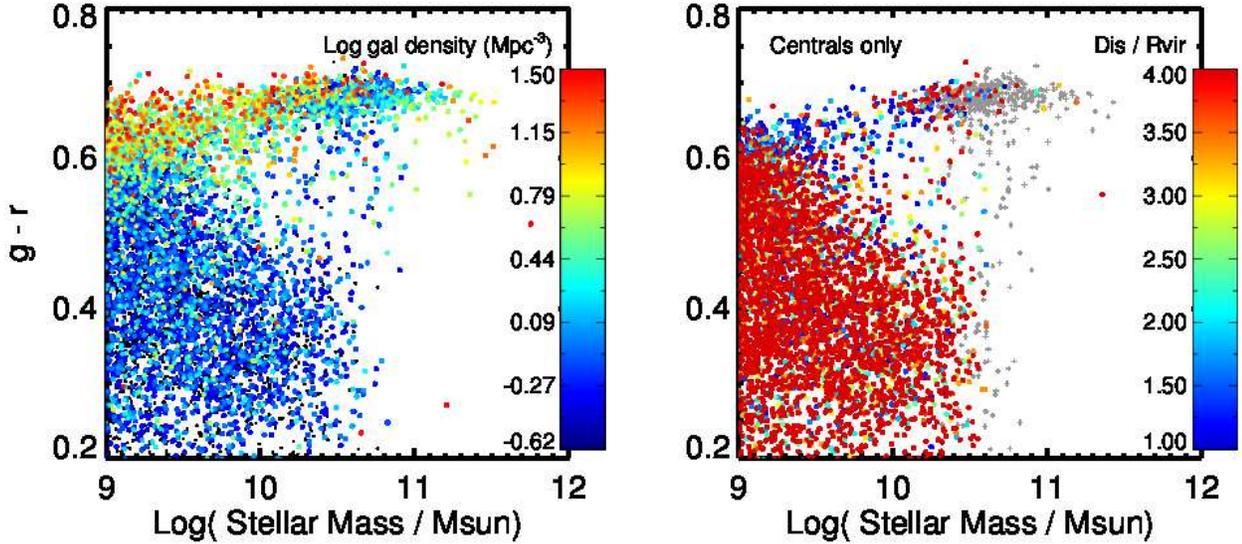}
\caption{ {\bf Left} CMD (at $z=0$) with galaxies colour-coded by the galaxy density within 1 Mpc.  The high- and low-mass ends of the
red sequence live in dense environments, whereas $\sim L^*$ red
sequence galaxies tend to live at lower densities.  {\bf Right}
Central galaxies are colour-coded by the distance to the closest
massive halo (right).  Massive halos are defined by a mass of $10^{12}
M_{\sun}$, as found using a spherical overdensity algorithm.  This is
roughly the mass at which a halo is dominated by hot gas ($>250,000$
K).  Grey crosses are galaxies at the centers of their (massive)
halos, and blue and red points are outside the virial radius of the
nearest massive halo.  Some passive galaxies $<10^{10.5}M_{\sun}$ are
quenched despite living outside $R_{\rm vir}$, suggesting the
influence of hot gas in unrelaxed super-group environments.  Quenched
galaxies at $\gtrsim 4 R_{\rm vir}$ live in halos just below $10^{12}
M_{\sun}$, where scatter in the $M_{\rm halo}-$hot gas relation
allows them to be hot-gas-dominated.}
\label{fig.density}
\end{figure*}

\subsection{Bivariate dependence of environment on colour and $M_{\rm stellar}$}
\label{sec.cmd_density}

We have shown that the massive and low-mass ends of the red sequence
grow first in our model -- not until $z\lesssim0.3$ do
intermediate-mass galaxies around $10^{10} M_{\sun}$ appear in
comparable numbers along the red sequence.  In this section, we
examine how environmental factors drive this behaviour.

We illustrate these environmental effects in Figure \ref{fig.density}.
On the left we show a colour-mass diagram where the galaxies are
colour-coded by their local galaxy density.  
The resulting trend is clear: the very highest and low-mass red
galaxies live in the densest regions, while intermediate-mass passive
galaxies ($M_*\sim 10^{10}-10^{11} M_{\sun}$) live in lower-density
regions.  This trend matches that seen in SDSS observations, at
least qualitatively \citep{hogg03, kauffmann04, blanton05_environ}.
Physically, this trend emerges because massive galaxies with $M_*\ga
10^{10.5} M_{\sun}$, once massive enough, can get to the red sequence
``on their own'' by forming their own hot halos even in (relatively)
low-density environments.  Massive galaxies that are quenched at
early epochs live in the highest overdensities, and thus grow to
be the most massive galaxies at $z=0$ by accreting other galaxies
(see \S \ref{sec.growth}).  Low-mass ($\lesssim 10^{10.5} M_{\sun}$)
galaxies can typically only be quenched by living in the hot
environment of more massive neighbors.

In the right panel of Figure \ref{fig.density} we show only {\it
central} galaxies colour-coded by the normalised distance to the nearest halo
with a mass $>10^{12} M_{\sun}$.  This is roughly the halo mass
where hot halos will form independently in our simulations.  Gray
crosses are for galaxies at the centers of
massive halos, and coloured points are for central galaxies in
lower-mass halos.

\begin{figure*} 
\includegraphics[width=84mm]{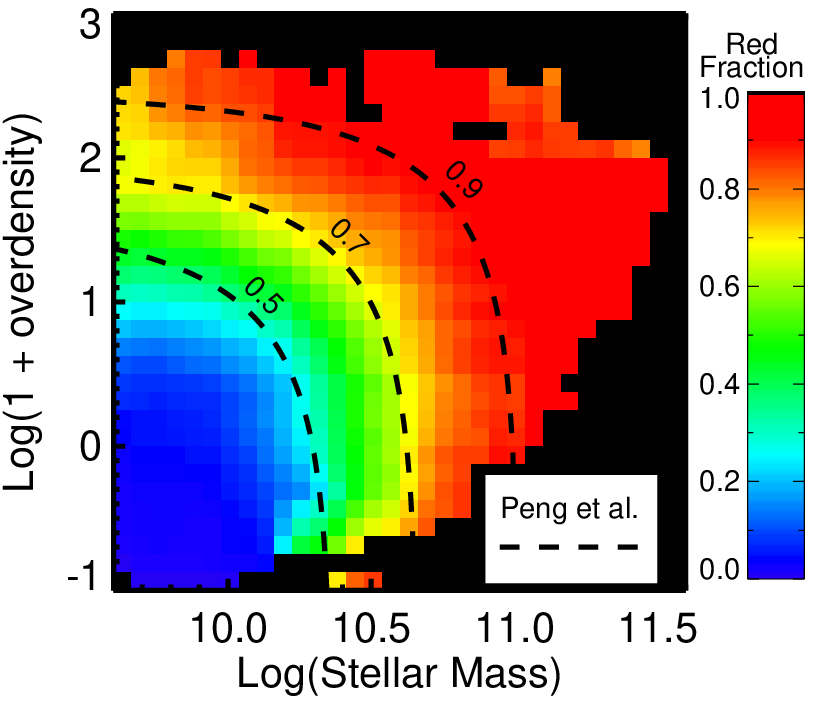}
\hfil
\includegraphics[width=84mm]{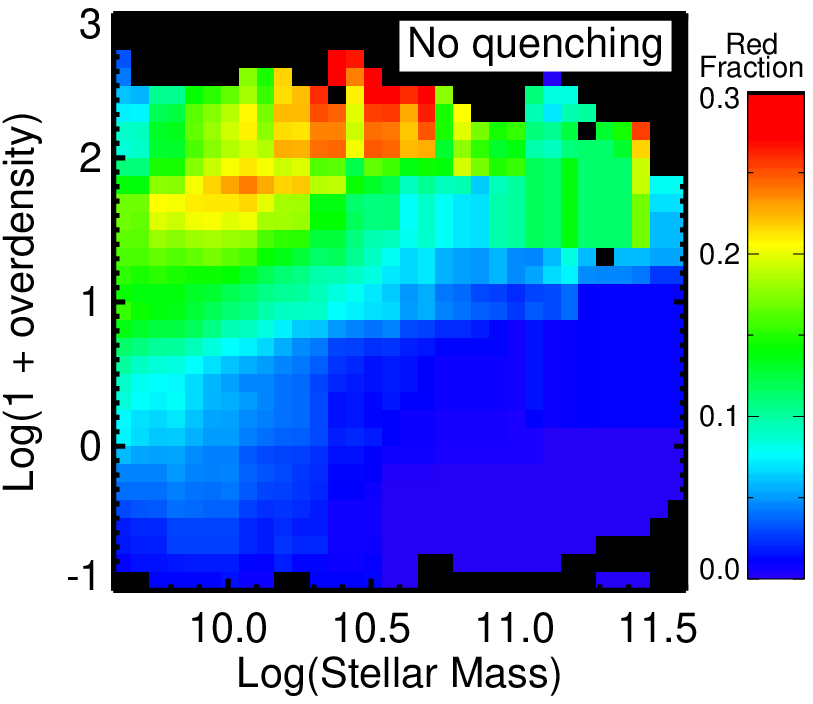}
 \caption{{\bf Left:} Smoothed distribution of red fraction (colors)
   in stellar mass vs. overdensity space in our quenching simulation.
   Galaxies are increasingly red at both higher masses and
   overdensities.  Small galaxies at high overdensity are red because
   they live within or close to larger halos dominated by hot gas,
   while larger galaxies are red because they are typically centrals
   living within hot halos maintained hot via our quenching feedback
   mechanism.  Dashed lines show the best-fit parameterization of red
   fraction of SDSS data from \citet{peng10}.  {\bf Right:} Same as
   left panel, but for a simulation without explicit quenching, and
   with a different color-scale.  Even without adding heat ``by hand''
   in hot halos, some satellite galaxies in high-density regions are
   quenched. }
\label{fig.fred_overdens}
\end{figure*}

\begin{figure*}
\includegraphics[width=168mm]{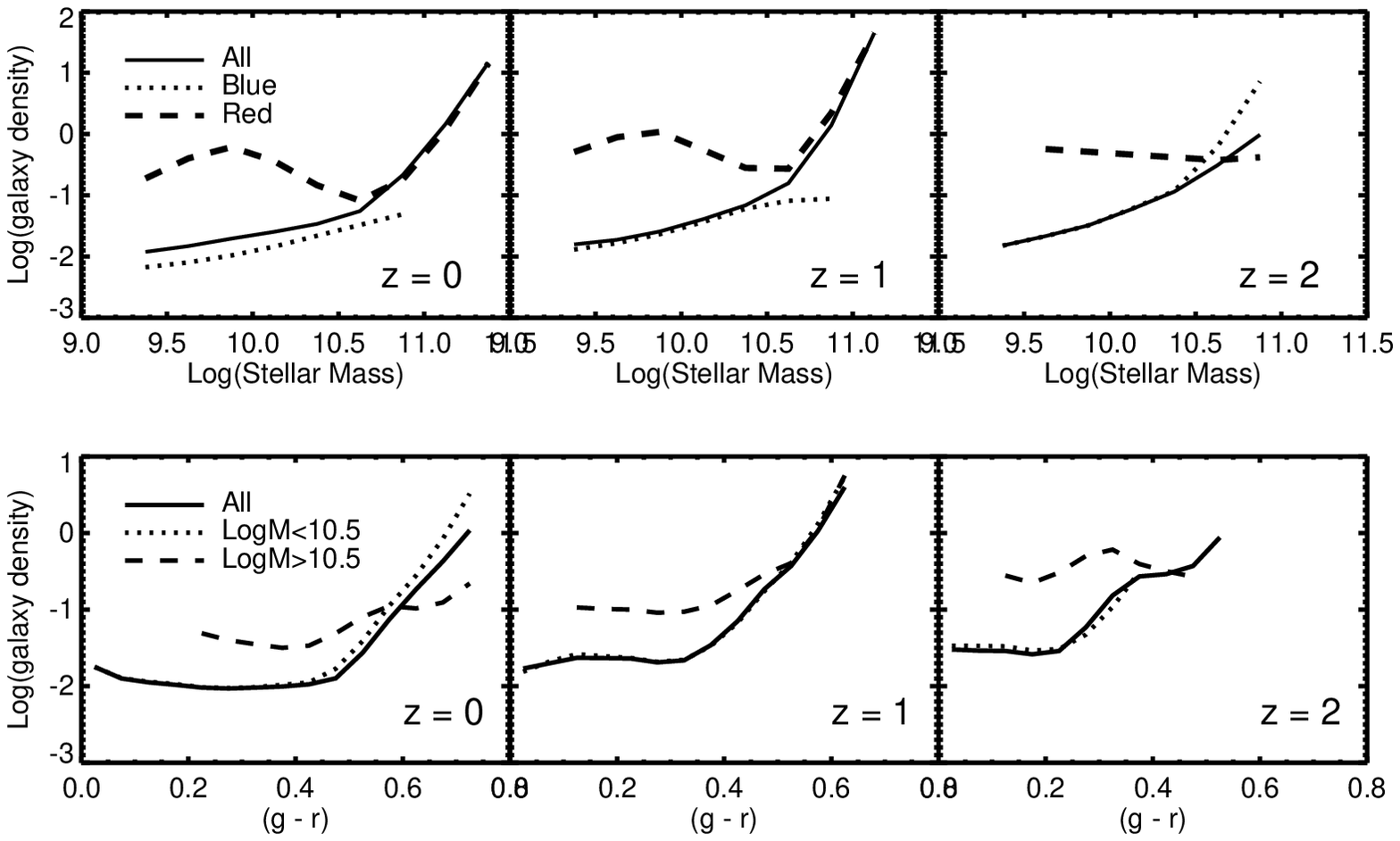}
\caption{{\bf Top row:} Mass-density relation for all (solid), blue
  (dotted), and red (dashed) galaxies at $z=0, 1,$ and 2 (increasing
  left-to-right). On average (in the ``all'' sample), galaxies below
  $10^{10.5} M_{\sun}$ live in low density environments, with a sharp
  rise in local density at higher masses.  \emph{Red} galaxies at low
  masses live in high-density environments, producing a pronounced dip
  in the red galaxy mass-density relation at $\sim 10^{10.5}M_{\sun}$.
  Redshift evolution to $z=1$ is negligible, and weak at higher $z$.
  {\bf Bottom row:} Colour-density relation for all (solid), low-mass
  (dotted), and high-mass (dashed) galaxies at three redshifts.  On
  average, blue galaxies live in low-density environments and red
  galaxies in high-density environments.  While the overall
  colour-evolution of galaxies is apparent in the different panels,
  the basic trends appear roughly the same at all $z$.  }
\label{fig.density_relations}
\end{figure*}

The first clear difference when looking only at centrals is that
there is a deficit of $M_*\sim 10^{10}M_\odot$ red central galaxies, which
as we argued in \S\ref{sec.centralsat} arises because red galaxies at that mass
are predominantly \emph{satellites} quenched at late times.

However, there are still some central galaxies that are along the red
sequence at intermediate to low masses.  Such galaxies at low mass
tend to be fairly close to a quenched halo (i.e. the points are blue).
This suggests a larger environmental effect -- these galaxies are part
of super-group structures whose hot gas can be felt beyond the virial
radius of the group halo.  Thus environmental factors beyond the halo
mass may play some role in quenching of star formation.  Observations
have suggested a role for environment (beyond halo mass) in some
galaxy properties \citep{cooper08, cooper08_MZR, cooper10}, though it
is typically secondary to galaxy mass.

Figure~\ref{fig.fred_overdens} shows another representation of the
relationship between red fraction and environment.  Here we plot
contours of red fraction (in colors from blue to red) in the plane of
stellar mass versus local galaxy overdensity (determined using {\tt
  Smooth}).  This figure can be compared to Figure~6 of
\citet{peng10}, who argued for two different quenching mechanisms,
namely ``environment quenching" of all galaxies above a given
overdensity, and ``mass quenching" above a given stellar mass.

Our simulation with quenching (left panel) quantitatively reproduces the observed trends almost
exactly, although it has far fewer galaxies than in the SDSS data of
\citet{peng10}, and hence the trends are less smooth and there are
regions in this space where this simulation yields no galaxies
(black).  Broadly, our models reproduce the strong increase in red
fraction to both high mass and high overdensity, and the ``boxy"
contour shape noted by \citet{peng10}.  In our simulations,
environment quenching is predominantly associated with satellites at
lower masses living in hot halos, although as discussed above there
are a fair number of small centrals living in dense environments that
are also quenched.  Meanwhile, mass quenching is well reproduced by
our hot gas threshold for quenching, which translates roughly into a
halo mass threshold.  To a large extent ``mass'' and ``environment'' quenching
  can alternatively be separated into central and satellite
  quenching \citep{delucia11_env, peng11}.

In the right panel of Figure~\ref{fig.fred_overdens} we
  illustrate ``environment quenching'' in our simulation without
  explicit quenching feedback.  This simulation produces fewer red
  galaxies \citep{gabor11}, and the figure shows that red galaxies
  form only at high overdensities.  The red galaxies in this case are
  almost all satellite galaxies.  ``Mass quenching,'' indicated by
  vertical contours in the left panel, is a direct result of adding
  heat to hot halos in our quenching simulation -- it is absent from
  the right panel.  Our explicit quenching mechanism also boosts the red
  fraction among satellites in dense regions.

In summary, our simulations are consistent with the interpretation by
\citet{peng10} that there are two quenching mechanisms that deliver
galaxies to the red sequence, associated with environment and mass,
though they can be described more physically as central and satellite
quenching.  But our simulations further suggest that both mechanisms
are ultimately the result of red galaxies living in regions dominated
by hot gas, which occurs at both high overdensities and high masses.

\subsection{Evolution of the mass-density and colour-density relations}
\label{sec.density_relations}

We show the evolution of the stellar mass--density relation and the
colour--density relation in Figure \ref{fig.density_relations}.  Here
the local galaxy density is found using {\tt Smooth}.

The mass-density relation (top row of Figure
\ref{fig.density_relations}) for all galaxies shows a very shallow
rise in density with mass for stellar masses below $10^{10.5}
M_{\sun}$.  Above this mass, density rises steeply.  This behavior
agrees qualitatively with observations at $z=0$ \citep{hogg03,
  blanton05_environ}.  The red galaxy sample departs from the average
trend because low-mass red galaxies live in dense regions.  For these
trends, evolution is negligible out to $z=1$ (top middle panel), but
notable at the massive end to $z=2$.  This high-$z$ evolution is sensitive to
the small numbers of massive galaxies at this redshift, and to the
separation between blue and red galaxies, so we do not view it as a
strong prediction at this time.

In the colour-density relation (bottom row of Figure
\ref{fig.density_relations}), density is uniformly low for blue
galaxies (low values of $g-r$), then rises steeply for red galaxies
(higher values of $g-r$).  Massive galaxies ($M>10^{10.5}M_{\sun}$,
dashed line) live at higher densities with a weaker colour-density
trend.  Low-mass galaxies, in contrast, \emph{must} live in dense
environments in order to become red.  Evolution in the colour-density
relation is minor to $z=1$, in accord with recent observations
\citep[although there is some debate][]{cooper06, cooper07, elbaz07,
  scodeggio09, cooper10}.

In summary, although our quenching model is most directly tied to hot
gas fraction and thus halo mass, environment plays a correlated role
in quenching galaxies in our simulations.  Halo mass is correlated
with environment in a CDM universe, so we get the usual mass-density
relation.  Since our quenching is directly linked to the proportion of
hot gas in a halo and thus halo mass \citep{birnboim03, keres05,
  gabor11}, we also naturally get a pronounced colour-density
relation.  Evolution in the mass-density and colour-density relations
is weak in this model and qualitatively consistent with data,
perhaps showing an inversion in the colour-density relation only at
the highest redshifts considered here.


\section{Summary and Conclusion} 
\label{sec.summary} 

We have analysed a cosmological hydrodynamic simulation to explore
the evolution of the red sequence.  A key feature of our simulation
is a simple quenching prescription tied to the dominance of hot gas
within the virial radius of a halo.  In our quenching prescription,
galaxies whose halos have $>60\%$ of their baryons in hot gas are
starved of star-forming fuel by continual heating of the halo gas.
In a previous paper we showed that this simple model yields a
distinct red sequence of galaxies whose number densities generally
agree with observations \citep{gabor11}.  This model thus provides
a tool for studying the emergence and growth of the red sequence
over cosmic time.  While some details of our model are overly
simplistic and in some cases are in conflict with observations, it
serves as representative of a general class of quenching models
where cessation of star formation is driven by the presence of hot
coronae or is tied to halo mass.

In our simulations, the first galaxies to turn red at $z\gtrsim 2$ are
among the most massive in the universe at that epoch, with $M_{\rm
  stellar}\sim 10^{10.5-11}M_{\sun}$.  Satellite galaxies at low
masses are concurrently starved of fuel by the same hot halo that
triggers our quenching mechanism, and therefore low mass galaxies also
appear on the red sequence at early times.  Meanwhile,
intermediate-mass galaxies at $\sim 10^{10}-10^{10.5} M_{\sun}$ remain
predominantly blue until late epochs, producing a pronounced dip in
the red galaxy mass function at high redshifts.  This dearth of
intermediate-mass red galaxies is qualitatively consistent with observations, but our model may have too many red galaxies with lower masses.  At late
times, the number of intermediate-mass red galaxies catches up to the
number of more massive ones, so that the dip in the red galaxy mass
function is less pronounced by $z=0$.

Essentially all simulated passive galaxies with stellar mass
$>10^{11}M_{\sun}$ form from smaller passive galaxies that grow mass
via mergers once on the red sequence. Most of the mass growth of
individual passive galaxies is due to minor mergers, with major
mergers being rare and a sub-dominant component of mass growth.  Based
on a simple analytic model, these numerous minor mergers can lead to
size growth that is a factor of 2 or more times that of a galaxy's
corresponding mass growth.  This suggests that minor mergers dominate
the dramatic size evolution of massive galaxies, in accordance with
earlier work \citep{naab09, hopkins10}.

The simple quenching model explored here qualitatively reproduces
observed trends among colour, stellar mass, and environment at $z=0$.
In detail, it also reproduces the observed mass-density trend for the
red sequence: low-mass and high-mass passive galaxies live in the
densest regions, whereas galaxies in-between (around $M^*$) live in
less dense regions.  The mass-density and colour-density trends
persist to at least $z=1$.  Quenching occurs at both high density,
roughly independent of stellar mass, and high stellar mass, roughly
independent of density.  As argued from corresponding observations by
\citet{peng10}, this indicates two quenching processes associated with
environment and mass, which in our model both arise due to starvation
when the local environment is dominated by hot gas.

Our analysis explictly avoids any considerations of galaxy
morphologies.  A key question that we therefore do not address is, why
are most red and dead galaxies elliptical?  While some authors argue
that it owes to major mergers that can transform star-forming spirals
into elliptical galaxies~\citep{mihos96}, the simple cessation of star
formation at sufficiently early epochs typically results in
substantial later growth via dry (i.e. purely or almost purely
stellar) mergers.  In this scenario, galaxies would quench first when
entering a hot halo (or its immediate environment), become red
spirals~\citep{skibba09_zoo, vanderwel09, bundy10, masters10,
  vanderwel11}, and then undergo subsequent dry mergers that alter its
morphology \citep[e.g.][]{delucia07, delucia11_bulge, guo08,
  feldmann10, oser10, oser12}.  Our models indicate that larger
galaxies living in denser environs undergo more such growth, and so
would be expected to have dynamical signatures indicating more dry
merging such as boxier isophotes \citep{naab06, bournaud07}; these
trends are qualitatively consistent with observations.  Whether the
quantitative predictions of this model are in accord with detailed
observations remains to be seen.

More broadly, it appears that a quenching mechanism based on a
simple hot gas fraction threshold (which is well-correlated with a
halo mass threshold) is capable of reproducing a variety of
observations not only at the present epoch but also out to $z\sim
1$ and beyond.  However, there are significant failures, such as a
dearth of the most massive galaxies, and a too-sharp cutoff in the
blue galaxy mass function suggesting that both today and at high
redshifts in the real Universe, some galaxies are able to be
star-forming despite living in hot gas-dominated halos.  Hence this
model represents only a step towards understanding the physical
processes driving red galaxy evolution.  Comparisons to observations
particularly probing the intermediate-mass regime out to high
redshifts can place strong constraints on models such as these, and
highlight additional physical processes that may need to be considered.


 
 

\section*{Acknowledgments}
The authors acknowledge K. Finlator, N. Katz, A. Maller,
B. D. Oppenheimer, and D. Weinberg for helpful discussions, the
referee for expeditious comments, and V. Springel for making Gadget-2
public. The simulations used here were run on the University of
Arizona's SGI cluster, ice. This work was supported by the National
Science Foundation under grant numbers AST-0847667 and AST-0907998,
and by the CANDELS Hubble Multi-Cycle Treasury program. Computing
resources were obtained through grant number DMS-0619881 from the
National Science Foundation.

\bibliographystyle{mn2e} 

\bibliography{paper}


\label{lastpage}

\end{document}